# High Deuterium Abundance in a New Quasar Absorber

Martin Rugers and Craig J. Hogan

University of Washington

Astronomy, Box 351580, Seattle WA 98195-1580

astro-ph/9603084    15 Mar 1996

## ABSTRACT

We present a new analysis of an absorption system in the spectrum of Q0014+813, with particular attention to a single cold cloud at redshift $z = 2.797957$. Features are identified at this redshift corresponding to SiIII, SiIV, CIV, DI, and HI, all consistent with purely thermal broadening at $T \approx 19,000$K. The deuterium identification is confirmed by its narrow width and precise agreement with the silicon and carbon redshifts. The HI column is well constrained by a slightly damped profile in the blue wing of Ly$\alpha$, and the DI column by its saturation, leading to a 95% confidence lower limit $(D/H) > 0.7 \times 10^{-4}$. The abundance is measured from line fits to be $(D/H) = 1.9^{+1.6}_{-0.9} \times 10^{-4}$, in agreement with the high deuterium abundance previously found in the $z = 3.32$ system in the same quasar spectrum. This system strengthens the case for a high universal primordial deuterium abundance and low cosmic baryon density ($\eta = 1.7 \times 10^{-10}$, $\Omega_b h^2 = 0.006$), for which Standard Big Bang Nucleosynthesis gives concordant predictions for helium-4 and lithium-7 abundances. Implications for Galactic chemical evolution and the baryonic and nonbaryonic dark matter problems are briefly discussed.

*Subject headings:* cosmology: observations — quasars: individual (0014+813)

## 1. Introduction

No place in the universe after the era of Big Bang nucleosynthesis is well suited to the manufacture and survival of deuterium, so even imprecise indications of ubiquitous cosmic deuterium confirm the standard picture of the early universe (Wagoner 1973, Reeves et al. 1973, Epstein et al. 1976). Indeed deuterium is largely destroyed even in the early Big Bang if the baryon density is high, so an accurate estimate of the primordial deuterium abundance $(D/H)_p$ gives a remarkably precise estimate of the cosmic baryon density. Precise knowledge of the primordial deuterium abundance is therefore central to understanding baryonic and nonbaryonic dark matter, and the related arguments concerning galaxy formation and clustering.

The primordial abundance is tricky to estimate because deuterium is so easily destroyed in stars; estimates of $(D/H)_p$ currently vary over almost almost an order of magnitude. This situation is rapidly changing however as high resolution, high signal-to-noise quasar spectra are now providing reliable, accurate estimates of $(D/H)_p$ from Lyman series absorption by HI and DI in foreground gas (Jenkins 1996). With access to a wide variety of lines, absolute deuterium abundances in some fortunately situated patches of high redshift gas can be estimated with reliability comparable to those in the local Galactic interstellar medium. Equally important (and unlike local measurements), we can sample systems with a wide range of chemical histories, including some with nearly primeval composition, where the primordial deuterium has been little affected by stellar destruction. This can remove or control much of the ambiguity in interpretation caused by chemical processing.

In this paper we present an analysis of a new high redshift quasar absorber of fairly pristine composition which provides useful constraints on deuterium abundance. The results confirm our earlier estimate from another absorber, indicating a universal high deuterium abundance. Comparison with other light element abundances supports the Standard

Big Bang Nucleosynthesis model and favors a low cosmic density of baryons, with broad implications for cosmological theory.

Songaila et al. (1994) and Carswell et al. (1994) showed that an absorber at $z = 3.32$ in Q0014+813 displays lines of $DI$ and $HI$ which permit accurate estimates of column densities of each, and therefore a reliable abundance of about $D/H = 2 \times 10^{-4}$, unless the deuterium line is caused by a chance hydrogen interloper. Rugers and Hogan (1996a) showed that this possibility is very unlikely, since the deuterium feature is actually caused by two narrow lines, each consistent with the width expected for deuterium, but too narrow to be an interloper from the Ly$\alpha$ forest; they derive an abundance $D/H = 1.9 \pm 0.5 \times 10^{-4}$ in both components. This result is in excellent concordance with the predictions of the standard Big Bang model and the other light element abundances as well as the known number of baryons in the universe (Hogan 1996), but it is surprising (and controversial) because it is significantly higher than previous upper limits based on solar system helium-3 measurements (Copi et al 1995a).

Other systems have been studied recently consistent with this result, but most of them have errors large enough to accomodate much smaller abundances. We have found $\log(D/H) \approx -3.95 \pm 0.54$ in the $z = 2.89040$ cloud in GC0636+680 (Rugers and Hogan 1996b), but here the errors do not provide a compelling lower limit. Songaila and Cowie (1996) find $\log(D/H) \approx -3.7$ in an absorber in Q 0956+122. Wampler et al. (1996) analyze a system at $z = 4.672$ in QSO BR 1202-0725. They find a high oxygen abundance, possibly indicating significant stellar processing, with an upper limit or possible detection of $(D/H) = 1.5 \times 10^{-4}$. Carswell et al. (1996) analyze a system at $z = 3.08$ in QSO 0420-388, and find an estimate of $\log(D/H) = -3.9 \pm 0.4$, or $-3.7 \pm 0.1$ if constant $OI/HI$ is assumed for the different redshift components, but can only provide a firm lower limit of $(D/H) > 2 \times 10^{-5}$. The reason for the ambiguity is that the column densities are difficult to measure accurately from highly saturated line profiles with multiple components. Similar systems have been discussed which give an apparently much lower abundance: $(D/H) = 2.3 \pm 0.3 \pm 0.3 \times 10^{-5}$ (at $z = 3.572$ in Q 1937-1009, Tytler et al. 1996), and $(D/H) = 2.4 \pm 0.3 \pm 0.3 \times 10^{-5}$ (at $z = 2.504$ in Q 1009+2956, Burles and Tytler 1996). Unless the high values are all spurious detections, these are best explained by local destruction of deuterium (see below). But as these authors point out, the case for a high deuterium abundance has rested squarely on just the $z = 3.32$ absorber of Q0014.

We present here evidence from the other well known Lyman limit system in Q0014. It is dominated by a single, simple, clean, high-column absorber with a large HI column, cold cloud on the blue side of the absorption complex. The interpretation is aided by metal line identifications, which lead to a precise determination of the gas redshift. These properties make it good target for estimating $(D/H)$ in spite of the fact that its higher order Lyman lines are not accessible (because of the Lyman limit absorption of the $z = 3.32$ system). The column density is high enough to detect absorption in the blue damping wing of Ly$\alpha$, which well constrains the HI column density once the redshift of the gas is fixed using silicon and carbon lines. The deuterium shows up as extra absorption which adds a sharp absorption edge to the blue wing of the damped HI profile. This closely constrains both its Doppler parameter and redshift. The metal lines imply very little turbulent broadening in this system, and indeed the fitted Doppler parameters of the HI and DI features are measured to be consistent with those expected in a purely thermal model, $b = \sqrt{kT/m}$, where $m$ is the mass of the absorbing species. The deuterium feature is too narrow to be caused by a Ly$\alpha$ forest interloper, and its saturation provides a solid lower limit to $D/H$, supporting a high primordial abundance.

## 2. Observations, Reductions, and Line Fits

We use the same data described in SCHR: a stack of six exposures of 40 minutes each of Q0014+813, using the HIRES echelle spectrograph on the Keck telescope, with resolution of 36,000. Reductions are as described in Rugers and Hogan (1996a). Redshifts quoted here are in vacuo, relative to the local standard of rest.

As in Rugers and Hogan (1996a), fits to the spectrum are performed using VPFIT, which models the formation of an absorption spectrum caused by a series of thermally broadened components at discrete velocities (Carswell et al. 1987, Webb 1987). For each component VPFIT determines the redshift ($z$), Doppler parameter ($b$) and column density ($N$) by simultaneously fitting the data to the summed opacity of all of the components, with the Voigt profiles convolved with a Gaussian instrument profile. It makes formal error estimates from the covariance matrix parameters, for each of the calculated $z$, $b$ and $N$, based on derivatives of the reduced $\chi^2$. In setting limits on parameters such as column density, we also use VPFIT to measure the $\chi^2$ directly to derive confidence levels.

The absorption complex near $z = 2.8$ is best fit by the components with parameters presented in Tables 1 and 2. We focus attention here on just one system with lines identified with HI, DI, CIV, SiIII, and SiIV, with redshifts near 2.797957. The components of interest are numbers 1, 8 and 9 in Table 1, and numbers 1 and 9 in Table 2. Figure 1 shows plots of the data and the simultaneous fit to the Ly-$\alpha$ complex and SiIV features. Figure 2 shows the fit to the SiIII features, plotted on the same velocity scale; these represent a separate fit since there is an unknown amount of Ly$\alpha$ forest contamination for many of the components (although not for the target system, since it is so narrow). A closeup of the CIV fit is shown in Figure 3. The characteristics of the fitted lines for the $z = 2.797957$ system are summarized in table 3.

The metal lines provide the key constraint on this system. There can be little doubt that they are metal lines, based on their narrowness alone. The SiIV identifications are confirmed by the relative optical depth and wavelength of the two lines, and the SiIII and CIV by agreement in redshift and width with SiIV.

The SiIV redshifts are used to fix the HI redshifts, in particular that of the most blueward large column density component of HI. The HI column density of this component is then well constrained by the blue damping wing of Ly$\alpha$. The damping profile is a good fit in the extreme blue wing, but requires another component to fit the sharp blue edge of the feature. Just one component is needed to fit this feature very well, and the best fit corresponds to DI at very nearly the same redshift as the HI, CIV and Si lines, to within $-0.16 \pm 1.2$ km/sec. The good fit to the blue side profile with only the single HI and DI components justifies the use of redshift from the silicon features.

The Lyman $\alpha$ line edge is so steep that the deuterium feature has a tightly constrained width, $b = 12.5 \pm 2.5$ km/sec, as illustrated in Figure 4. This is too narrow to be a hydrogen interloper; out of 262 lines identified by Hu et al. (1995) in a 600Å portion of this spectrum, only three have $b \leq 14$.[1] Using this as a guide to the density of lines which might be contaminants, we find that the Poisson probability of an unidentified metal line occuring in the $1\sigma$ redshift range of the DI feature $\delta z = \pm 0.000015$ (a total wavelength interval of 0.036Å) by chance is about $1.8 \times 10^{-4}$. The deuterium identification is much more plausible.

The unresolved metal lines indicate a very small turbulent broadening for this system, so we can use the deuterium feature to measure the temperature; $b = 12.5 \pm 2.5$ km/sec gives $T = 1.9 \pm 0.7 \times 10^4$K. The predictions for the thermal widths of the other features are then in good agreement with those observed (see table 3), confirming this interpretation and the deuterium line identification.

---

[1] Two of these have positive metal identifications, one of which is identified here as SiIII. Since they are metal lines, correlations with HI at the same wavelength are not expected, justifying the use of Poisson statistics in calculating interloper probability. Note that our SiIV lines lie outside the spectral range of the Hu et al. study.

As a consistency check, we can get a very rough idea of the physical state of the cloud using Donahue and Shull's (1991) models. (Since the cloud is optically thick at the Lyman edge and the models do not include radiative transfer, the ionizing spectrum is harder in the cloud than in the models. This may lead to errors of up to factors of a few in $U$ and $Z$, but not more than about 10% in $T$; see Madau 1991). The SiIII/SiIV ratio indicates an ionization parameter of about $\log U = -3$, corresponding to an HI neutral fraction of about 0.01; the $SiIV/HI$ ratio then gives an abundance from the models of about 0.01 times solar. The predicted equilibrium temperature is $T = 1.7 \times 10^4$K, in agreement with that estimated from the widths of the lines. The ionization parameter is small enough ($U \leq 10^{-2}$) for charge-exchange reactions $D^+ + H \rightleftharpoons D + H^+$ to guarantee that $H$ and $D$ are locked to the same fractional ionization, so $N_{DI}/N_{HI}$ indeed gives the true absolute deuterium abundance of the gas.

The deuterium abundance from the line fits is $\log(D/H) = \log N_{DI} - \log N_{HI} = -3.73 \pm 0.28$. The central value, corresponding to $(D/H) = 1.9 \times 10^{-4}$, agrees with that found for the other Lyman limit cloud in Q0014 (Rugers and Hogan 1996a). The errors here are somewhat larger ($(D/H) = 1.9^{+1.6}_{-0.9} \times 10^{-4}$), mainly because the D feature is saturated. However, the lower limit from this system, $(D/H) \geq 1 \times 10^{-4}$, is more solid than these errors might suggest. The damping wing gives a secure upper limit to the HI column, and the lower limit to the DI column is guaranteed by the saturation. We have evaluated this limit from the reduced $\chi^2$ of the fit over the 2Å near the line edge, which increases steeply for low assumed $N_{DI}$: while $(D/H)$ below $1 \times 10^{-4}$ is ruled out at the 84% level, $(D/H)$ below $0.7 \times 10^{-4}$ is ruled out at least at the 95% level, and $(D/H)$ below $0.5 \times 10^{-4}$ at the 99% level. This saturation constraint is illustrated in Figure 4, which shows the best fitting model with $(D/H) = 0.7 \times 10^{-4}$, comparing it to the data and the best fit. The constraint on the high side is not as tight; even an abundance as high as $10^{-3}$ is ruled out only at the 90% level.

## 3. Discussion

We have identified features in two of Q0014's high redshift Lyman limit systems as deuterium lines. The redshift and width of the DI features in both systems agree with those predicted for DI at the same velocity and temperature as the hydrogen and other species. Such narrow lines are rare in quasar spectra, making the probability of chance interlopers at the precise required velocities very small, much less than $10^{-3}$ for each case. The HI and DI column in both cases are well constrained once the identification is accepted, leading to reliable abundance determinations. The new case studied here makes the case for high $(D/H)$ much more compelling, even though the $z = 3.32$ system still allows a more precise measurement, $(D/H) = 1.9 \pm 0.5 \times 10^{-4}$.

A fascinating and suggestive coincidence is that the well-measured clouds in this quasar— that is, the $z = 2.8$ cloud and both components of the $z = 3.32$ cloud— all give the same deuterium abundance. This result, together with the fact that no higher abundances have been found, may be an observational hint that we are seeing a universal primordial abundance. While the two cloud systems lie in the same direction, they are separated by an enormous distance, $\Delta z = 0.52$. This result may confirm the fine grained uniformity of the early universe at widely separated places.

A high primordial deuterium abundance conflicts with earlier estimates, based on measurements of helium-3 in the solar system together with educated guesses about the amount of destruction of deuterium and its further processing in stars over the history of the Galaxy (Hata et al 1995, Copi et al 1995ab). However, even with sophisticated and detailed modeling of chemical evolution these estimates depend critically and quantitatively on uncalibrated assumptions about stellar processing, especially about how thoroughly stellar populations process the destroyed deuterium beyond helium-3. The quasar technique gives a more direct and reliable estimate of $(D/H)_p$, while the local data convey information primarily on Galactic chemical history. The present result supports independent theoretical and observational arguments (Hogan 1995, Charbonnel 1995, Wasserburg et al 1995, Weiss et al 1996) that many stars destroy helium-3 efficiently before ejecting their envelopes.

Variations among different quasar absorbers could similarly be explained by localized destruction of deuterium, which would initially be very patchy. It is useful to note that the total mass of gas actually covering the quasar image and doing the absorption is very small, perhaps $10^{-6}$ solar masses in a typical case, and that even over 1 pc or so from the line of sight the absorber contains only about one stellar mass, hence is easily influenced by ejecta from just a single stellar envelope. Therefore while widespread deuterium destruction must be accompanied by metal production, this is only true statistically and does not necessarily apply in detail; especially in such metal-poor systems, and over such a small sampling volume, the processing is likely to be due to a very small number of stars, which could very well not include any stars sufficiently massive to eject metals. We thus consider it premature to draw conclusions about primordial spatial variations in $\eta$ from just a handful of even very metal-poor systems. The best way to estimate the primordial abundance is to obtain a sample of systems with reliable determinations and seek a "ceiling" to the observed abundances, which may now be emerging from the data.

If the primordial deuterium abundance indeed is $(D/H)_p = 1.9 \times 10^{-4}$, how does this fit with cosmological theory? Standard Big Bang Nucleosynthesis (Copi et al. 1995a, Sarkar 1996) gives this abundance for a baryon-to-photon ratio $\eta = 1.7 \times 10^{-10}$, corresponding to a density parameter in baryons $\Omega_b h^2 = 0.006$.[2] For this value of $\eta$, the primordial helium abundance is predicted to be $Y_p = 0.233$, in excellent agreement with current estimates from HII regions in nearby metal-poor galaxies, for example $Y_p = 0.228 \pm 0.005 \pm 0.005$ (Pagel et al. 1992), $Y_p = 0.231 \pm 0.006$ (Skillman and Kennicutt 1993, Skillman et al. 1993), $Y_p = 0.229 \pm 0.004$ (Izotov et al. 1994), and a compilation including systematic error, $Y_p = 0.232 \pm 0.003 \pm 0.005$ (Olive and Steigman 1995). The primordial lithium abundance is predicted to be $\log(Li/H) = -9.75 \pm 0.2$ (theoretical error only), in excellent agreement with the value estimated from metal-poor halo stars, for example $\log(Li/H) = -9.80 \pm 0.16$ (compiled by Walker et al. 1991, Smith et al. 1993), $\log(Li/H) = -9.78 \pm 0.20$ (Thorburn 1994). The Big Bang model therefore works well with this relatively high deuterium abundance and correspondingly low baryon density. Another expression of this agreement is the nucleosynthesis estimate of the number of families of light particles, which agrees with the standard model $N_\nu = 3$ (Cardall and Fuller 1996, Sarkar 1996); Peimbert (1996) obtains $N_\nu \approx 3.1 \pm 0.5$.

The density parameter in baryons $\Omega_b h^2 = 0.006$, or $\Omega_b = 0.012$ for $h = 0.7$, also fits with the mean cosmic density of known baryonic systems. Stars and gas in known galaxy populations contribute about $\Omega \approx 0.005$ (Fukugita et al. 1996), X-ray emitting intergalactic gas in groups and clusters adds $\Omega \approx 0.001$ (Persic and Salucci 1992, Salucci and Persic 1996), and cool intergalactic absorbers at the present epoch add about $\Omega \approx 0.001$, for a total density in known baryons of about $\Omega = 0.007$. Thus even with $\Omega_b = 0.012$ we are faced with many uncounted baryons, but only about as many as those already observed (Dar 1995). These may be in the form of new galaxy populations (Bristow and Phillips 1994), compact baryonic objects (Alcock et al. 1995), or ionized intergalactic gas. Indeed, the ratio of global baryonic mass density to global stellar blue light density is about $10h^{-1}$ in solar units, which applied to our Galaxy (assuming it is typical) implies a total baryonic component of about $1.7 \times 10^{11} h^{-1}$ solar masses. This is consistent with recent estimates of the mass of baryonic MACHOs in the Galactic halo from gravitational microlensing (Alcock et al. 1995, 1996).

---

[2] If we take the errors from the $z = 3.32$ system, $(D/H)_p = 1.9 \pm 0.5 \times 10^{-4}$, the density estimate including errors is $\eta = 1.7 \pm 0.3 \times 10^{-10}$ or $\Omega_b h^2 = 0.006 \pm 0.001$. Of course, a larger uncertainty than this attaches to the assumption that this system indeed gives the primordial abundance: absorption by interloping HI always introduces a slight upward bias, and deuterium destruction a downward bias, both of which can be large in individual cases.

Of course a much larger global density of *nonbaryonic* dark matter is still required, especially to satisfy galaxy cluster and group dynamics. Since typical galaxies which dominate the luminosity density of the universe indeed tend to have mass-to-blue-light ratios exceeding $10h^{-1}$ in solar units ($30h$ is more typical; see Rubin 1993), we conclude that nonbaryonic dark matter is ubiquitous in the universe, even in galactic halos. At the same time, guesses at the total global density parameter $\Omega$ in nonbaryonic dark matter, based on assuming that the baryon mass per total mass in galaxy clusters reflects the global ratio (White et al 1993), typically decrease as a consequence of low $\Omega_b$.

We are particularly grateful to A. Songaila and L. L. Cowie for performing the observations and sharing the data, and to R. Carswell for use of his VPFIT software and for help in using it. This work was supported at the University of Washington by NSF grant AST 932 0045 and NASA grant NAG-5-2793.

## REFERENCES

Alcock, C., et al. 1995 *Phys. Rev. Lett.* **74**, 2867

Alcock, C., et al. 1996, preprint.

Bristow, P. D. and Phillips, S. 1995, *MNRAS* **267**, 13

Burles, S. and Tytler, D. 1996, *Science*, submitted; astro-ph 9603070

Cardall, C. Y. and Fuller, G. M. 1996, astro-ph 9603071

Carswell, R.F., Webb, J.K., Baldwin, J.A., & Atwood, B. (1987), *ApJ*, **319**, 709

Carswell, R F., Rauch, M., Weymann, R. J., Cooke, A. J., and Webb, J. K., 1994, *MNRAS* **268**, L1

Carswell, R F., Webb, J. K., Baldwin, J.A., Cooke, A.J., Williger, G.M., Rauch, M., Irwin, M.J., Robertson, J.G. and Shaver, P.A. 1996, *MNRAS*, **278**, 506

Charbonnel, C. 1995, *ApJ*, **453**, L41.

Copi, C. J., Schramm, D. N., and Turner, M. S. 1995a, *Science* **267**, 192.

Copi, C. J., Schramm, D. N., and Turner, M. S. 1995b, *Phys. Rev. Lett.* **75**, 3981

Dar, A. 1995, *ApJ*, **449**, 550

Donahue, M. & Shull, J. M. 1991, *ApJ*, **383**, 511.

Epstein, R. I., Lattimer, J. M., and Schramm, D. N. 1976, Nature 263, 198

Fukugita, M., Hogan, C. J. and Peebles, P. J. E., 1996 *Nature*, in press.

Hata, N., Scherrer, R. J., Steigman, G., Thomas, D., Walker, T. P., Bludman, S., Langacker, P., 1995, *Phys. Rev. Lett.* **75**, 3977

Table 1.  VPFIT results for the absorbers in Figures 1.

|    | Ion  | logN  | ±    | $z$      | ±        | $b$ (km/s) | ±    |
|----|------|-------|------|----------|----------|------------|------|
| 1  | SiIV | 12.83 | 0.13 | 2.797957 | 0.000008 | 3.1        | 2.9  |
| 2  | SiIV | 12.63 | 0.12 | 2.798914 | 0.000095 | 34.3       | 10.9 |
| 3  | SiIV | 12.77 | 0.08 | 2.799461 | 0.000015 | 13.0       | 2.7  |
| 4  | SiIV | 13.22 | 0.11 | 2.799821 | 0.000005 | 4.8        | 1.2  |
| 5  | SiIV | 12.59 | 0.30 | 2.800242 | 0.000078 | 20.9       | 17.5 |
| 6  | SiIV | 12.78 | 0.18 | 2.800654 | 0.000029 | 13.0       | 4.7  |
| 7  | SiIV | 13.17 | 0.09 | 2.801108 | 0.000019 | 19.0       | 1.9  |
| 8  | DI   | 14.31 | 0.25 | 2.797955 | 0.000015 | 12.5       | 2.5  |
| 9  | HI   | 18.04 | 0.12 | 2.797957 | -        | 19.2       | 5.6  |
| 10 | HI   | 15.60 | 1.46 | 2.798914 | -        | 46.8       | 12.4 |
| 11 | HI   | 16.04 | 2.10 | 2.799416 | -        | 63.3       | 44.2 |
| 12 | HI   | 17.16 | 2.93 | 2.799821 | -        | 53.5       | 30.1 |
| 13 | HI   | 16.44 | 2.29 | 2.800242 | -        | 33.5       | 26.3 |
| 14 | HI   | 16.59 | 0.57 | 2.800654 | -        | 27.3       | 22.7 |
| 15 | HI   | 17.08 | 0.39 | 2.801108 | -        | 19.1       | 7.5  |
| 16 | HI   | 12.86 | 0.09 | 2.802757 | 0.000053 | 25.5       | 6.4  |
| 17 | HI   | 13.09 | 0.05 | 2.803722 | 0.000026 | 20.9       | 3.2  |
| 18 | HI   | 13.07 | 0.08 | 2.805078 | 0.000105 | 57.2       | 12.5 |
| 19 | HI   | 13.97 | 0.03 | 2.806896 | 0.000008 | 20.8       | 1.1  |
| 20 | HI   | 13.49 | 0.09 | 2.807585 | 0.000120 | 54.8       | 10.2 |

Table 2. VPFIT results for the absorbers in Figures 2 and 3.

|   | Ion | logN | ± | $z$ | ± | $b$ (km/s) | ± |
|---|---|---|---|---|---|---|---|
| 1 | SiIII | 13.24 | 0.22 | 2.797955 | 0.000009 | 3.8 | 3.1 |
| 2 | SiIII | 12.58 | 0.07 | 2.798941 | 0.000006 | 5.2 | 1.9 |
| 3 | SiIII | 13.10 | 0.19 | 2.799494 | 0.000017 | 18.8 | 3.4 |
| 4 | SiIII | 13.36 | 0.14 | 2.799939 | 0.000100 | 41.1 | 9.8 |
| 5 | SiIII | 13.47 | 0.07 | 2.800320 | 0.000070 | 59.2 | 6.1 |
| 6 | HI | 13.15 | 0.04 | 2.768003 | 0.000017 | 18.1 | 2.2 |
| 7 | HI | 14.30 | 0.10 | 2.771553 | 0.000011 | 18.8 | 2.6 |
| 8 | HI | 13.51 | 0.03 | 2.773196 | 0.000022 | 31.1 | 2.3 |
| 9 | CIV | 12.95 | 0.11 | 2.797959 | 0.000007 | 3.2 | 2.7 |

– 18 –

Table 3. Overview of the line parameters for the $z = 2.797957$ cloud.

| Species | Line(s) | $v - v_0$[†](km/s) | $b$ (km/s) | $b_{Th}$ (km/s) | $\log N$ |
|---------|---------|-------------------|------------|-----------------|----------|
| CIV | 1548.2 Å | $0.16 \pm 0.55$ | $3.2 \pm 2.7$ | $5.1 \pm 1.0^*$ | $12.95 \pm 0.11$ |
| SiIV | 1393.8 Å, 1402.8 Å | $0.0 \pm 0.63$ | $3.1 \pm 2.9$ | $3.3 \pm 0.7^*$ | $12.83 \pm 0.13$ |
| SiIII | 1206.5 Å | $-0.16 \pm 0.71$ | $3.8 \pm 3.1$ | $3.3 \pm 0.7^*$ | $13.24 \pm 0.22$ |
| HI | Ly-$\alpha$ | $(\equiv 0.0 \pm 0.0)$ | $19.2 \pm 5.6$ | $17.7 \pm 3.5^*$ | $18.04 \pm 0.12$ |
| DI | Ly-$\alpha$ | $-0.16 \pm 1.18$ | $12.5 \pm 2.5$ | $\equiv 12.5 \pm 2.5$ | $14.31 \pm 0.25$ |

[†]Relative to $z = 2.797957$. The HI fit is fixed at this redshift. Errors are for relative velocity only; the absolute scale is uncertain by about 0.06Å or about 4 km/sec.

[*]Prediction for thermal widths if $b_{DI} = 12.5 \pm 2.5$ km/s, corresponding to $T = 19,000 \pm 7,000$ K.

## Figures

Figure 1. Spectral fits to Lyman $\alpha$, SiIV(1393.755 Å), and SiIV(1402.770 Å), plotted on the same velocity scale. Ticks mark the centroids of individual components, with parameters tabulated in Table 1; the overall model fit comprising the sum of this absorption is the solid line. Components 9 through 15 are constrained to have the same $z$ as 1 through 7 respectively. The important hydrogen line is component 9, defined to be at the same $z$ as component 1 in SiIV. The DI component (number 8) is found at very nearly the same $z$, and is responsible for the sharp blue edge of the line.

Figure 2. Spectral fit to SiIII(1206.510 Å) on the same velocity scale as Figure 1. The fit parameters are tabulated in Table 2. These components are not constrained to lie at the same $z$ as those in table 1, but component 1 closely corresponds in $z$ to the dominant cloud (components 1, 8, and 9 in table 1). The rest of the fit includes five components with approximate counterparts in table 1 (1,2,3, 5 here corresponding approximately to 1,2,3, 7 in table 1), and components 6 through 8, which are suspected Ly-$\alpha$ forest interlopers.

Figure 3. The fit to the CIV line, corresponding to the bluemost hydrogen line. Only the 1548.188Å line is measured, as the other CIV lines fall in the coverage gap between orders 37 and 38 in the echellogram. The fit parameters are listed as component 9 in Table 2.

Figure 4. Detail of two fits near the Ly$\alpha$ line edge. The two components 1 and 2 correspond to D and H Ly$\alpha$ respectively, both fixed at the SiIV redshift $z = 2.797957$. The best fit curve corresponds to $(D/H) = 1.9 \times 10^{-4}$ with Doppler width $b = 12.5$km/sec. The other theoretical curve corresponds to our 95% confidence lower limit, $(D/H) = 0.7 \times 10^{-4}$, for which the best fit Doppler parameter is $b = 19.04 \pm 1.32$ km/sec. This model clearly does not provide enough absorption to match the saturation in the data. Such large model Doppler parameters, which would be required to produce this feature with an HI interloper, do not produce a sufficiently steep line edge to match the data regardless of redshift.

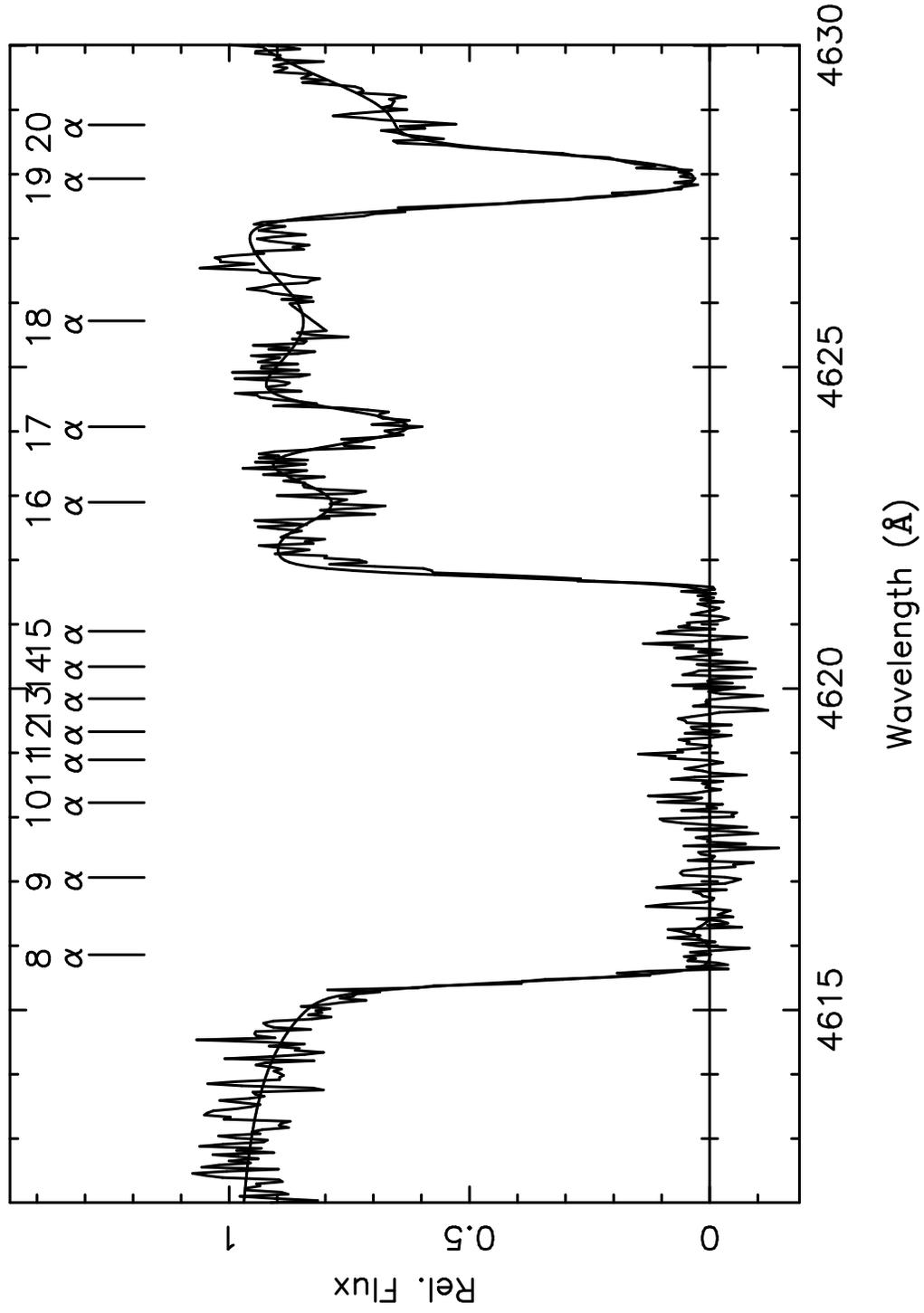

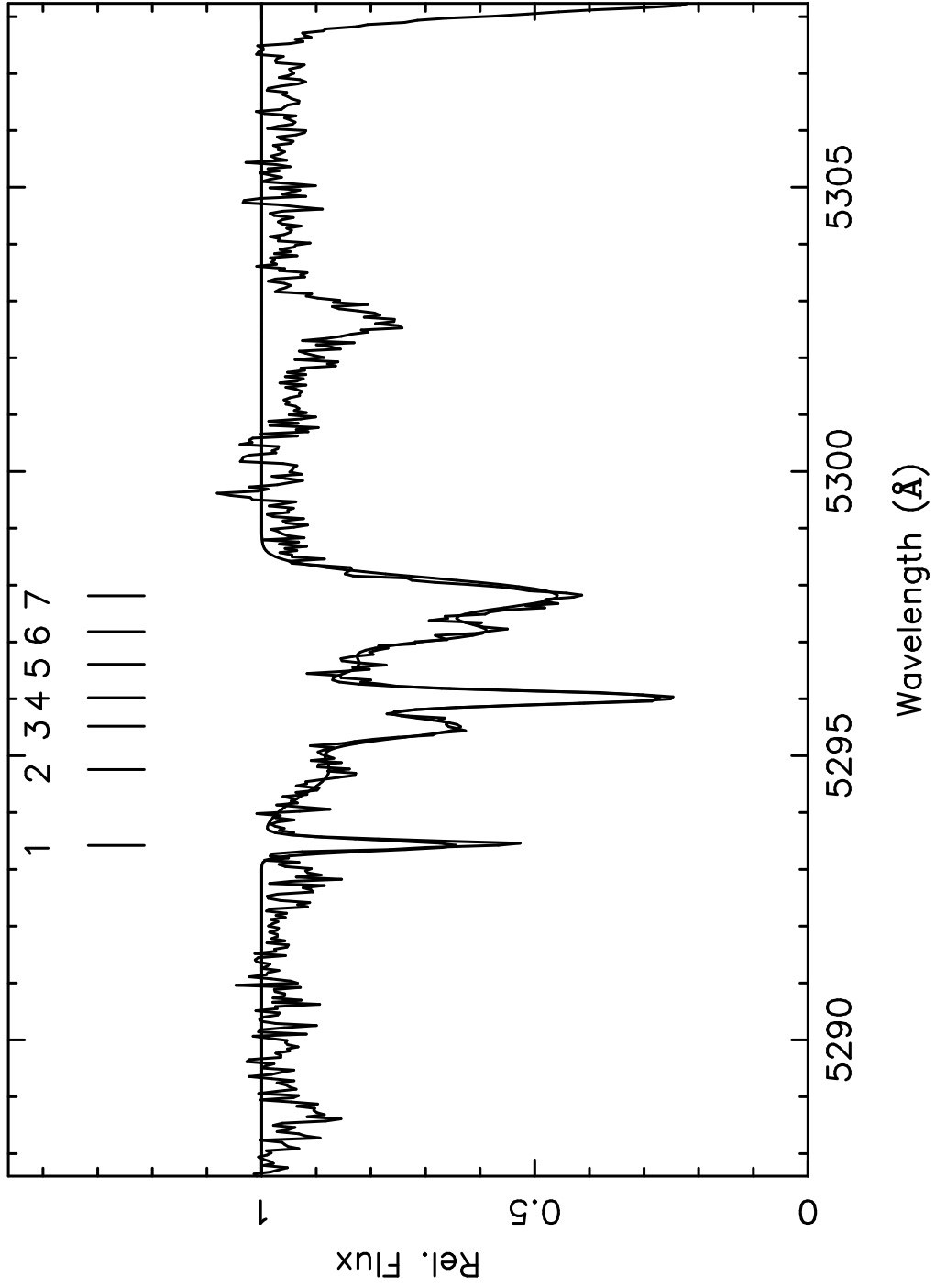

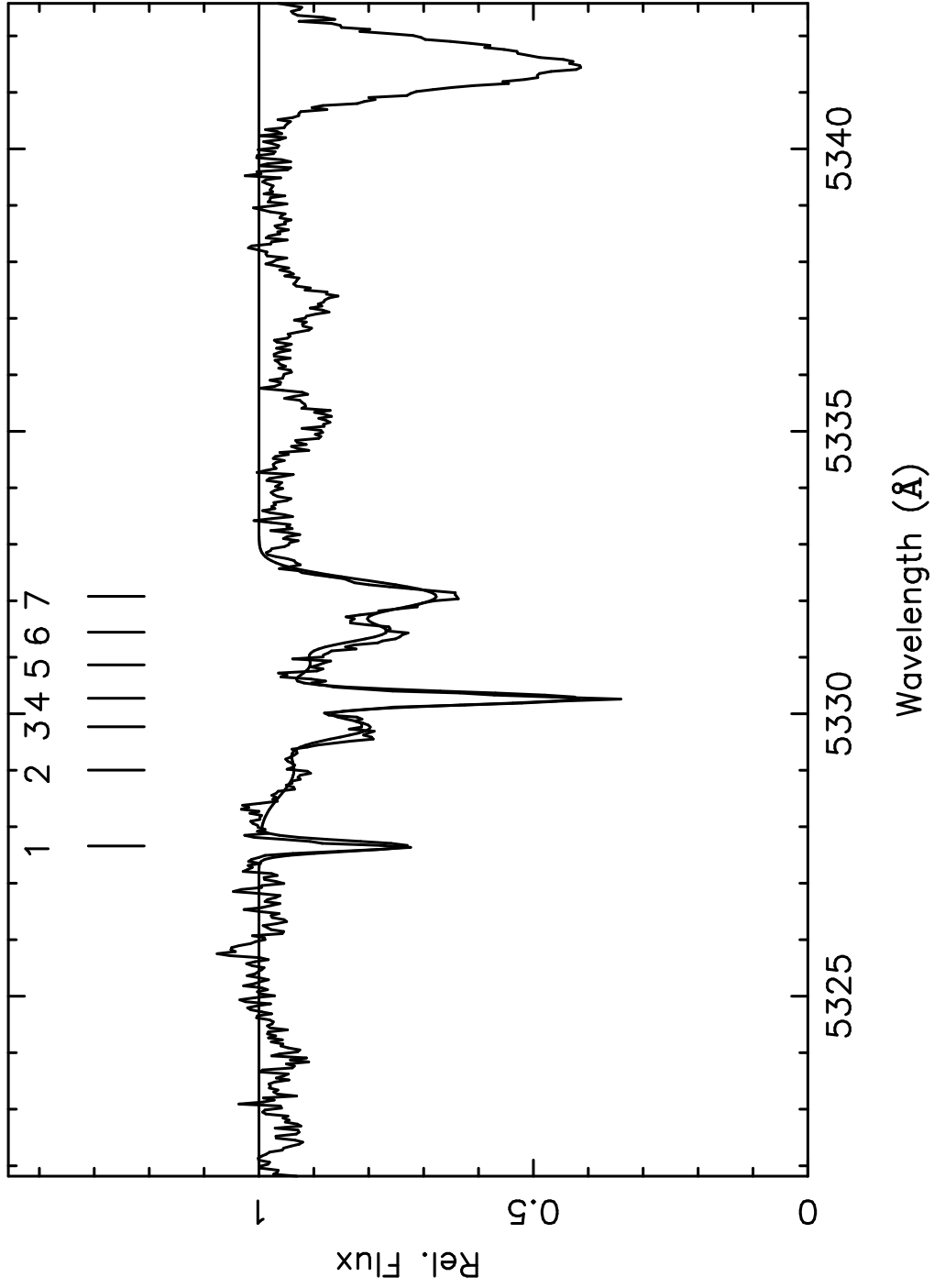

Fits for SiIV (1402.8 Å) of the z=2.80 LLS in Q0014.

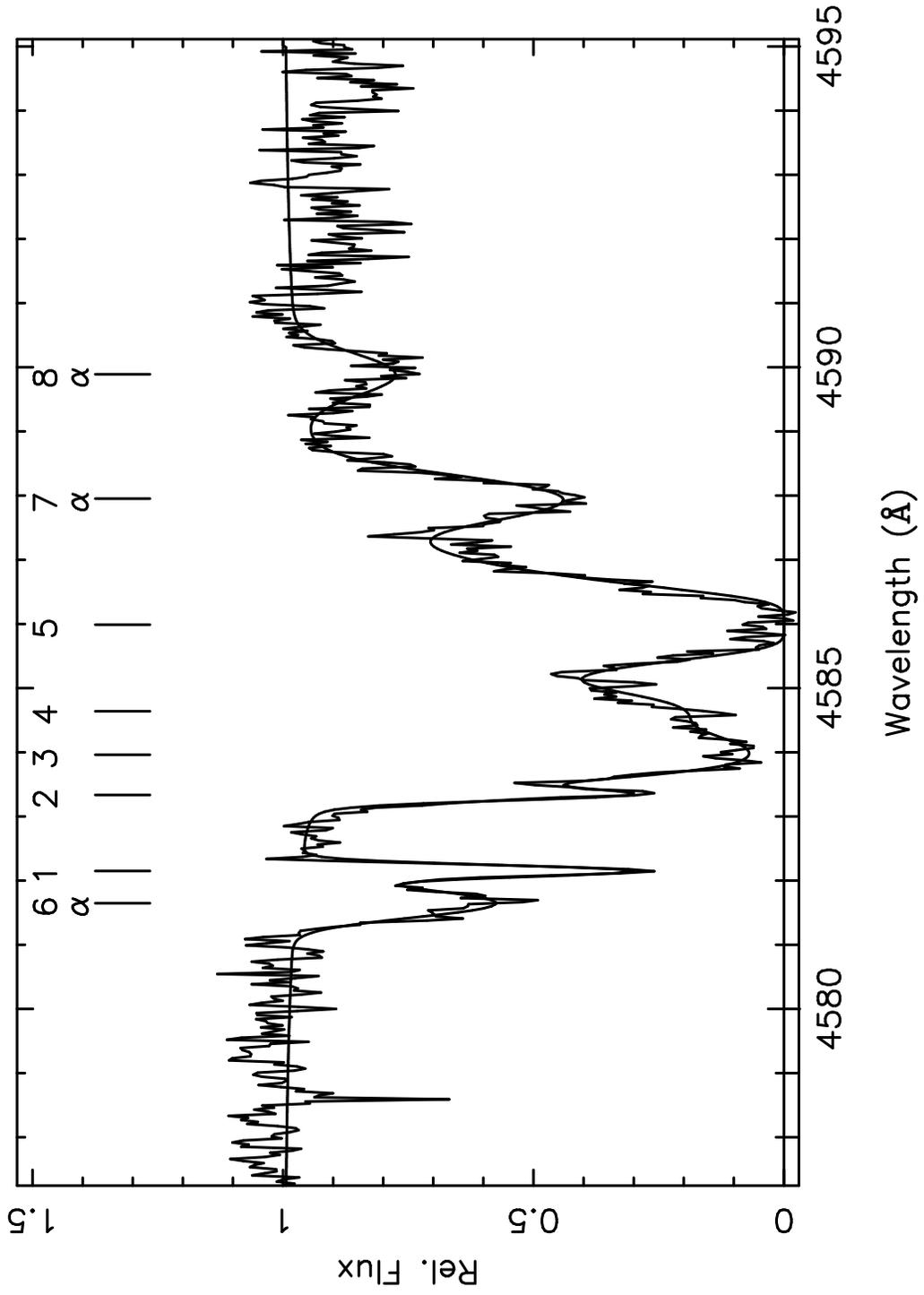

Fits for SiIII (1206.5 Å) and Ly-α interlopers.

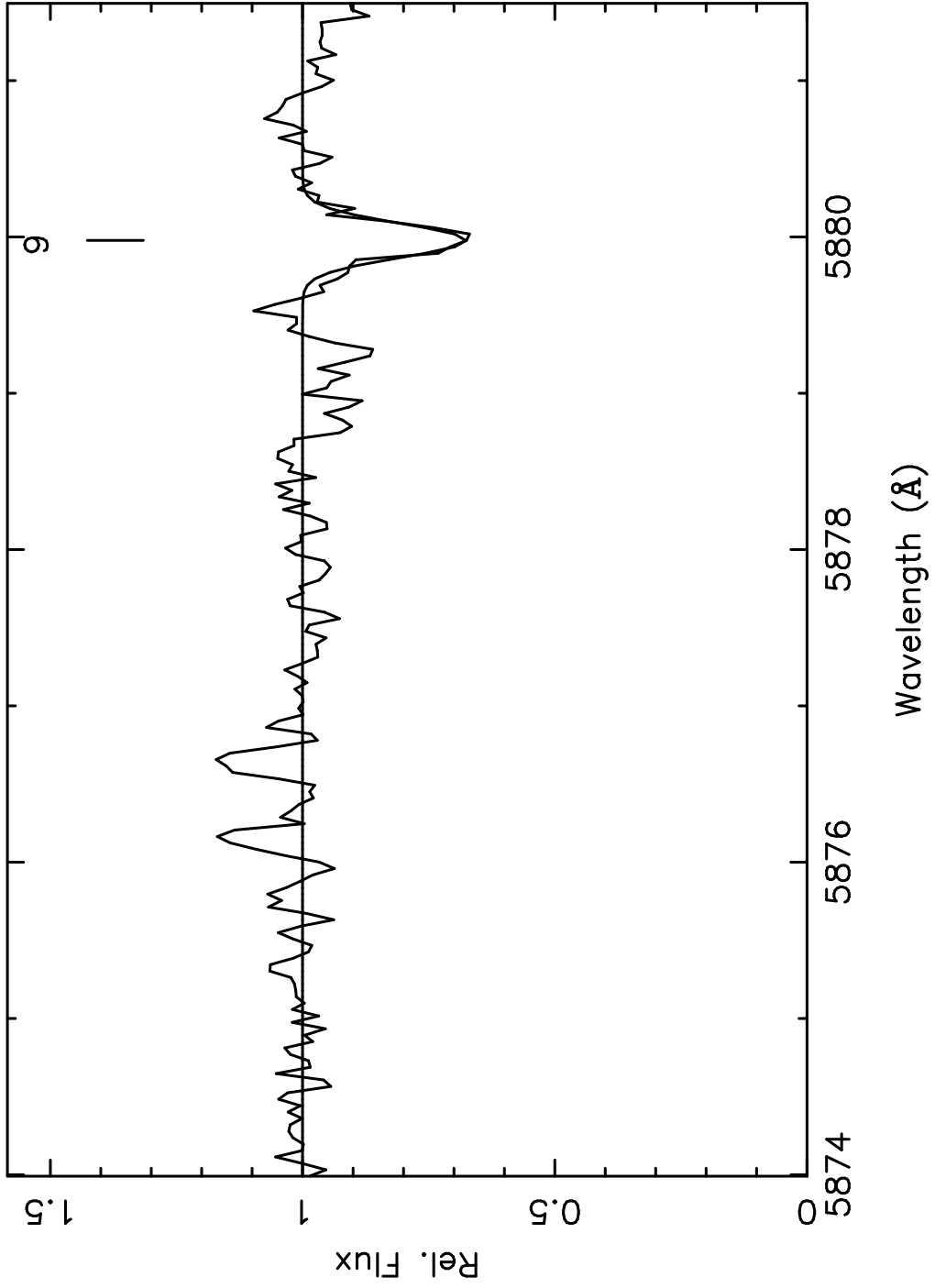

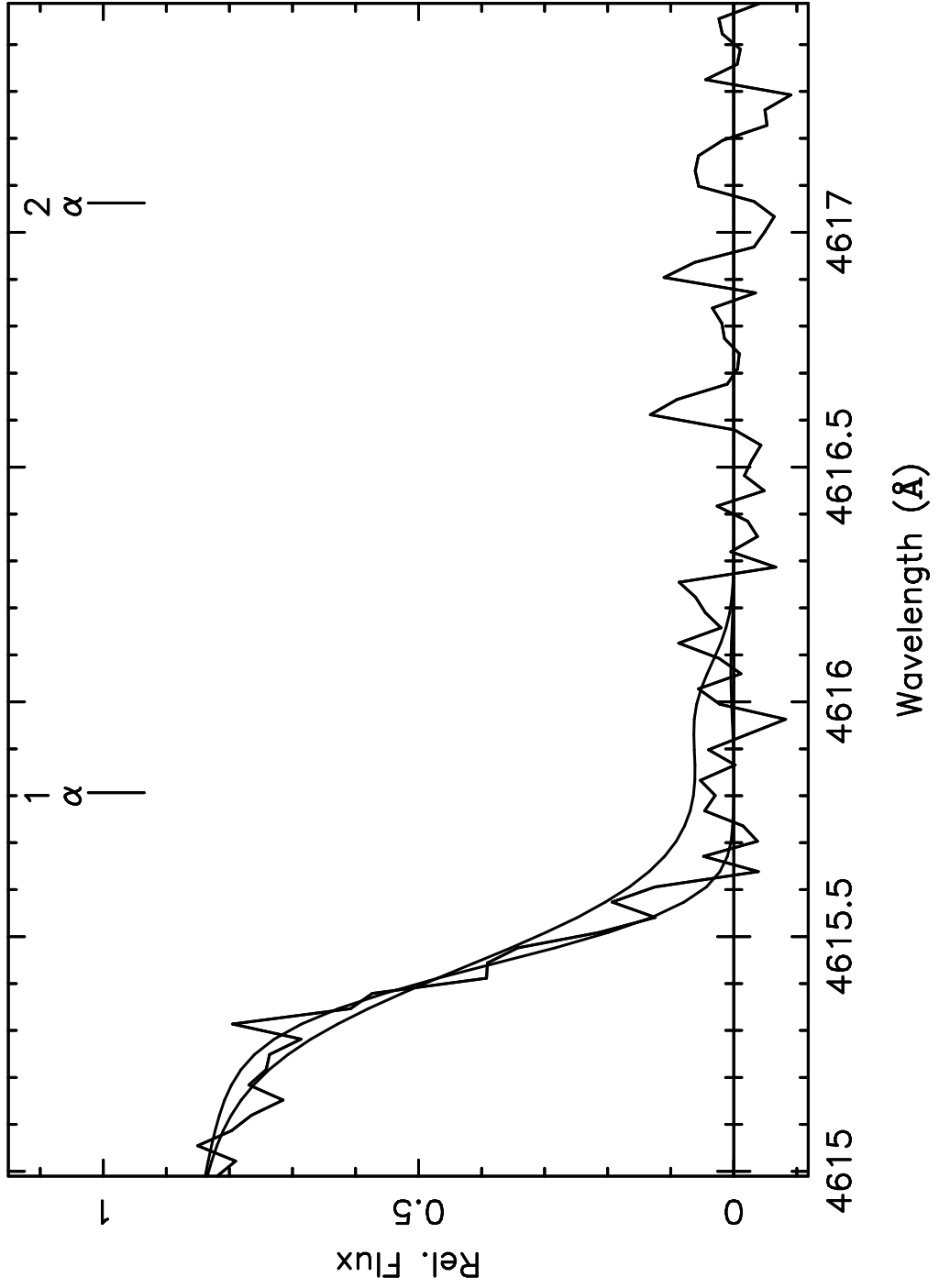